\documentclass{JHEP3}
\usepackage{graphicx}
\usepackage{amsfonts}
\usepackage{amssymb}
\usepackage{cite}

\title{Integrated Markov Chain Monte Carlo (MCMC) analysis of primordial non-Gaussianity ($f_{\mathrm{NL}}$) in the recent CMB data}
\author{Jaiseung Kim $^1$\\Niels Bohr Institute \& Discovery Center, Blegdamsvej 17, DK-2100 Copenhagen, Denmark\\$^1$\email{jkim@nbi.dk}}
\date{\today}

\abstract{
We have made a Markov Chain Monte Carlo (MCMC) analysis of primordial non-Gaussianity ($f_{\mathrm{NL}}$) using the WMAP bispectrum and power spectrum.
In our analysis, we have simultaneously constrained $f_{\mathrm{NL}}$ and cosmological parameters so that the uncertainties of cosmological parameters can properly propagate into the $f_{\mathrm{NL}}$ estimation.
Investigating the parameter likelihoods deduced from MCMC samples, we find slight deviation from Gaussian shape, which makes a Fisher matrix estimation less accurate. Therefore, we have estimated the confidence interval of $f_{\mathrm{NL}}$ by exploring the parameter likelihood without using the Fisher matrix. 
We find that the best-fit values of our analysis make a good agreement with other results, but the confidence interval is slightly different.
}
\keywords{CMBR experiments, CMBR theory, non-gaussianity}
\preprint{arXiv:1009.0981}
\makeindex

\begin{document}
\section{Introduction}
Over the past years, there have been great successes in measurement of the CMB anisotropy by ground and satellite observations  \cite{WMAP7:basic_result,ACBAR2008,QUaD2,Planck_bluebook,Planck_mission}.
CMB anisotropy, which is associated with the inhomogeneity of the last scattering surface, provides invaluable information on our Universe.
In particular, the angular power spectrum of CMB anisotropy provides strong constraints on cosmological models and cosmological parameters \cite{Modern_Cosmology,Inflation,Foundations_Cosmology}. 

Large classes of cosmological inflation models predict some level of primordial non-Gaussianity ($f_{\mathrm{NL}}$), whose detailed shape and amplitude depend on the particular class of the inflationary model \cite{Komatsu_thesis,fnl_Bartolo_review,Komatsu_fnl_review}.
Therefore, an investigation of the imprints of primordial non-Gaussianity allows us to constrain and rule out some classes of inflationary models \cite{fnl_power,fnl_Bartolo_review,Komatsu_fnl_review}.
Since the release of the COBE and subsequently WMAP data \cite{Mather:CMB_T_1999,Fixen:dipole,Bennett:dipole,WMAP1:Gaussianity,WMAP3:temperature,WMAP7:basic_result}, there have been active investigations on the imprints of primordial non-Gaussianity, using CMB data from the satellite and sub-orbital experiments 
\cite{bispectrum_cobe,WMAP1:Gaussianity,fnl_Yadav,fnl_WMAP3,fnl_subl,WMAP5:Cosmology,fnl_Smith,fnl_Archeops,fnl_boomerang,fnl_needlet,fnl_wavelet,WMAP7:Cosmology,fnl_wavelet7,Komatsu_fnl_review}. Currently, the strongest limit is imposed on the local-type $f_\mathrm{NL} = 32 \pm 21$ by the WMAP data \cite{WMAP7:Cosmology,Komatsu_fnl_review}. 
CMB bispectrum, which is most sensitive to $f_{\mathrm{NL}}$ parameters, also have some dependence on cosmological parameters \cite{Komatsu_thesis,fnl_Yadav,Komatsu_fnl_review}. 
Noting this, we have simultaneously constrained $f_{\mathrm{NL}}$ and cosmological parameters by using the WMAP bispectrum and power spectrum.
In this way, we have allowed the uncertainties of cosmological parameters to properly propagate to the $f_{\mathrm{NL}}$ estimation. 
The parameter likelihoods deduced from MCMC samples show slight deviation from Gaussian shape, which makes Fisher matrix estimation less accurate. 
Therefore, we have estimated the confidence intervals of $f_{\mathrm{NL}}$ by fully exploring the parameter likelihood without using Fisher matrix. 
We find our estimation makes a good agreement with the existing results, but the confidence interval of $f_{\mathrm{NL}}$ is slightly different.

The outline of this paper is as follows.
In Section \ref{primordial_NG}, we discuss the primordial non-Gaussianity.
In Section \ref{CMB} and \ref{ML}, we discuss the CMB bispectrum and its likelihood function.  
In Section \ref{analysis}, we make an analysis of WMAP data and present the results.
In Section \ref{Discussion}, we summarize our analysis and discuss prospects. 

\section{Primordial perturbations}
\label{primordial_NG}
Primordial perturbation in Fourier space is given by \cite{CMB_fnl,Komatsu_thesis,fnl_simulation,WMAP5:Cosmology}:
\begin{eqnarray}
\Phi(\mathbf k)=\Phi_{\mathrm L}(\mathbf k)+\Phi_{\mathrm {NL}}(\mathbf k),\label{Phi_Fourier}
\end{eqnarray}
where $\Phi_{\mathrm L}(\mathbf k)$ and $\Phi_{\mathrm {NL}}(\mathbf k)$ denote the primordial perturbation of purely Gaussian nature and the deviation respectively. 
In most of inflationary models, 
$\Phi_{\mathrm L}(\mathbf k)$, which follows a Gaussian distribution and statistical isotropy, have the following statistical properties \cite{Modern_Cosmology,Inflation,Foundations_Cosmology}:
\begin{eqnarray}
\langle \Phi_{\mathrm L}(\mathbf k_1) \Phi_{\mathrm L}(\mathbf k_2) \rangle &=&(2\pi)^3 P_{\Phi}(k)\,\delta(\mathbf k_1+\mathbf k_2),\label{Phi_L_2point}
\end{eqnarray}
where $\langle \ldots \rangle$ denotes the average over an ensemble of universes, and $P_{\Phi}(k)=A\,k^{n-4}$ is a primordial power spectrum with a normalization amplitude $A$. The non-Gaussian part of the primordial perturbation, $\Phi_{\mathrm {NL}}(\mathbf k)$, has negligible 2-point correlation \cite{fnl_bias,fnl_power}, which is also severely constrained by observational data: 
\begin{eqnarray}
\langle \Phi_{\mathrm {NL}}(\mathbf k_1) \Phi_{\mathrm {NL}}(\mathbf k_2) \rangle &\approx&0.\label{Phi_NL_2point}
\end{eqnarray}
Additionally, we may consider a 3-point correlation.
For Gaussian $\Phi_{\mathrm {L}}(\mathbf k)$, we expect a vanishing 3-point correlation \cite{Modern_Cosmology,Inflation,Foundations_Cosmology,Komatsu_thesis}:
\begin{eqnarray}
\langle \Phi_{\mathrm {L}}(\mathbf k_1)\,\Phi_{\mathrm {L}}(\mathbf k_2)\,\Phi_{\mathrm {L}}(\mathbf k_3)\rangle=0 \label{Phi_L_3point}.
\end{eqnarray}
On the other hand, a non-vanishing 3-point correlation of $\Phi_{\mathrm {NL}}(\mathbf k)$ is predicted by large classes of inflationary models:
\begin{eqnarray}
\langle \Phi_{\mathrm {NL}}(\mathbf k_1)\,\Phi_{\mathrm {NL}}(\mathbf k_2)\,\Phi_{\mathrm {NL}}(\mathbf k_3)\rangle=(2\pi)^3 \delta(\mathbf k_1+\mathbf k_2+\mathbf k_3)\,F(k_1,k_2,k_3) \label{Phi_NL_3point}.
\end{eqnarray}
where $F(k_1,k_2,k_3)$ denotes a primordial bispectrum, whose detailed shape depends on the specific inflationary model \cite{fnl_Bartolo_review}.
In this paper, we shall consider the ``local'' and ``equilateral'' forms of the bispectrum, which are respectively associated with the squeezed and equilateral configuration of three vectors \cite{CMB_fnl,Komatsu_thesis,fnl_simulation,fnl_Bartolo_review,Maldacena_fnl,WMAP5:Cosmology,fnl_orth,WMAP7:Cosmology,Komatsu_fnl_review}.
The primordial bispectrum in its ``local'' and ``equilateral'' forms is defined as follows \cite{fnl_Bartolo_review,Yadav_fnl_review,Komatsu_fnl_review}:
\begin{eqnarray}
F_{\mathrm{local}}(k_1,k_2,k_3)&=&2 A^2 f^{\mathrm{local}}_{\mathrm{NL}} \left[\frac{1}{k^{4-n}_1 k^{4-n}_2}+\frac{1}{k^{4-n}_2 k^{4-n}_3}+\frac{1}{k^{4-n}_3 k^{4-n}_1}\right],\nonumber\\
F_{\mathrm{equil}}(k_1,k_2,k_3)&=&6 A^2 f^{\mathrm{equil}}_{\mathrm{NL}}\left[-\frac{1}{k^{4-n}_1 k^{4-n}_2}-\frac{1}{k^{4-n}_2 k^{4-n}_3}-\frac{1}{k^{4-n}_3 k^{4-n}_1}-\frac{2}{(k_1 k_2 k_3)^{2(4-n)/3}}\right.\nonumber\\
&&\left.+\left(\frac{1}{k^{(4-n)/3}_1 k^{2(4-n)/3}_2 k^{4-n}_3}+(\mathrm{5\:perm.})\right)\right],\nonumber
\end{eqnarray}
where $f^{\mathrm{local}}_{\mathrm{NL}}$ and $f^{\mathrm{equil}}_{\mathrm{NL}}$ denote the non-linear coupling parameters for ``local'' and ``equilateral'' forms respectively.
The ``local'' form possesses a special significance because
 all single-field inflation models predict a negligible $f^{\mathrm{local}}_{\mathrm{NL}}$. Therefore, the detection of a high  $f^{\mathrm{local}}_{\mathrm{NL}}$ would allow us immediately to rule out single-field inflation models \cite{singlefield_fnl_consistency,singlefield_fnl_general,singlefield_fnl,inflation_fieldtheory,Komatsu_fnl_review}.
Besides the aforementioned forms, there are also other forms for primordial bispectrum, which are predicted by certain inflation models \cite{fnl_excited_init,fnl_warm_inflation,fnl_quasi_singlefield,fnl_warm_inflation2}.
Therefore, the physical observables associated with the 3-point correlation provides important information on our early Universe.

\section{The CMB anisotropy}
\label{CMB}
The CMB anisotropy over a whole-sky is conveniently decomposed in terms of spherical harmonics: 
\begin{eqnarray*}
T(\hat{\mathbf n})&=&\sum_{lm} a_{lm}\,Y_{lm}(\hat{\mathbf n}),
\end{eqnarray*}
where $a_{lm}$ and $Y_{lm}(\hat {\mathbf k})$ are a decomposition coefficient and a spherical harmonic function. 
The decomposition coefficient $a_{lm}$ is related to the primordial perturbations $\Phi(\mathbf k)$ according to:
\begin{eqnarray}
a_{lm}&=&4\pi (-\imath)^l \int \frac{d^3\mathbf k}{(2\pi)^3} \Phi(\mathbf k)\,g_{l}(k)\,Y^*_{lm}(\hat {\mathbf k}),\label{alm}
\end{eqnarray}
where $g_{l}(k)$ is a radiation transfer function.
Largely, we may consider CMB power spectrum and bispectrum, which are associated with 2-point and 3-point correlation of primordial perturbation respectively.
In the absence of tensor perturbation, the expectation value of CMB power spectrum is given by:
\begin{eqnarray}
C_{l}&=&\langle a_{lm}\,a^*_{lm} \rangle \\
&=&\frac{2}{\pi} \int k^2 dk g^2_{l}(k)\,P_{\Phi}(k). \label{Cl}
\end{eqnarray}
As shown in Eq. \ref{Phi_NL_2point}, 2-point correlation of $\Phi_{\mathrm {NL}}(\mathbf k)$ is insignificant. Therefore, the CMB power spectrum is mainly associated with $\Phi_{\mathrm {L}}(\mathbf k)$ \cite{fnl_bias,fnl_power} (c.f. Eq. \ref{Phi_L_2point}).
% For optimal estimators of CMB power spectrum,refer to \cite{Tegmark:likelihood,pseudo_Cl,MASTER,hybrid_estimation,WMAP3:temperature,WMAP5:powerspectra,WMAP7:powerspectra} for details.

Considering the rotational properties of spherical harmonics, we can construct a rotationally invariant estimator of the CMB bispectrum 
as follows \cite{CMB_momentum}:
\begin{eqnarray}
B^{\mathrm{obs}}_{l_1 l_2 l_3}=\sum_{m_1  m_2 m_3} \left(\begin{array}{ccc} l_1& l_2 & l_3 \\m_1& m_2& m_3\end{array}\right) 
 a_{l_1m_1} a_{l_2 m_2} a_{l_3 m_3}.\label{B_obs}
\end{eqnarray}
Hereafter, the condition ($l_1\le l_2 \le l_3$) will be implied, unless stated otherwise.
The expectation value of the CMB bispectrum, in other words, the theoretical CMB bispectrum, can be split into quantities associated with the different possible shapes of primordial non-Gaussianity:
\begin{eqnarray*}
\langle B^{\mathrm{obs}}_{l_1 l_2 l_3} \rangle = \sum_{\alpha} f^{i}_{\mathrm{NL}}\,B^{i}_{l_1 l_2 l_3},
\end{eqnarray*}
where $i$ denotes ``local'' and ``equilateral'' in our analysis.
The CMB bispectrum of $f^{i}_{\mathrm{NL}}=1$ is given by \cite{Komatsu_thesis,fnl_orth,WMAP5:Cosmology,WMAP7:Cosmology,Komatsu_fnl_review}:
\begin{eqnarray}
B^{\mathrm{local}}_{l_1 l_2 l_3}&=&2I_{l_1 l_2 l_3}\int^{\infty}_0 r^2 dr\left[\alpha_{l_1}(r)\beta_{l_2}(r)\beta_{l_3}(r)+\mathrm{2\:perm.}\right],\label{B_local}\\
B^{\mathrm{equil}}_{l_1 l_2 l_3}&=&-3B^{\mathrm{local}}_{l_1 l_2 l_3}+6\,I_{l_1 l_2 l_3}\int^{\infty}_0 r^2 dr \left[\beta_{l_1}(r)\gamma_{l_2}(r)\delta_{l_3}(r)+(\mathrm{5\:perm.})-2\delta_{l_1}(r)\delta_{l_2}(r)\delta_{l_3}(r)\right],\nonumber\\\label{B_equil}
\end{eqnarray}
where  ``perm.''  denotes cyclic permutation and
\begin{eqnarray}
I_{l_1 l_2 l_3}=\sqrt{\frac{(2l_1+1)(2l_2+1)(2l_3+1)}{4\pi}}\left(\begin{array}{ccc} l_1& l_2 & l_3 \\ 0& 0& 0\end{array}\right),\label{Illl}
\end{eqnarray}
\begin{eqnarray}
\alpha_l(r)&=&\frac{2}{\pi}\int k^2 dk\,g_{Tl}(k)\,j_l(kr),\label{alpha}\\
\beta_l(r)&=&\frac{2}{\pi}\int k^2 dk P_{\Phi}(k)\,g_{Tl}(k)\,j_l(kr),\label{beta}\\
\gamma_l(r)&=&\frac{2}{\pi}\int k^2 dk P^{1/3}_{\Phi}(k)\,g_{Tl}(k)\,j_l(kr),\label{gamma}\\
\delta_l(r)&=&\frac{2}{\pi}\int k^2 dk P^{2/3}_{\Phi}(k)\,g_{Tl}(k)\,j_l(kr),\label{delta}
\end{eqnarray}
with $j_l(x)$ being a spherical Bessel function.
In our analysis, we have computed Eq. \ref{B_local} and \ref{B_equil} by numerical integration.
We have investigated the functional shape of the integrand and used 10, 10, 100 and 30 integration points for the intervals  $r\le 6.5$, $6.5< r\le 13$,
$13<r\le 16.5$ and $16.5<r\le 39.5$ [Gpc] respectively. 
Compared with a much denser sampling ($\times 50$), we find that the difference in the estimation of $f^{\mathrm{local}}_{\mathrm{NL}}$ is less than $1\%$.
For $k$ space integration in Eq. \ref{alpha}, \ref{beta}, \ref{gamma} and \ref{delta}, we used the $k$ space sampling of \texttt{CAMB} \cite{CAMB}.
As discussed in Eq. \ref{Phi_L_3point}, 3-point correlation of $\Phi_{\mathrm {L}}(\mathbf k)$ is expected to be zero.
%there exist 3-point correlation only for $\Phi_{\mathrm {NL}}(\mathbf k)$.
Therefore, CMB bispectrum is not associated with $\Phi_{\mathrm {L}}(\mathbf k)$, but only with $\Phi_{\mathrm {NL}}(\mathbf k)$ (c.f. Eq. \ref{Phi_NL_3point}).

Due to foreground contamination, CMB anisotropy cannot be measured reliably over the whole sky  \cite{WMAP3:temperature,WMAP5:fg,WMAP7:fg}.
Given the spherical harmonic coefficients of a cut sky, we can construct the following estimator:
\begin{eqnarray}
\tilde B^{\mathrm{obs}}_{l_1 l_2 l_3}&=&\sum_{m_1 m_2 m_3}\left(\begin{array}{ccc} l_1& l_2 & l_3 \\m_1& m_2& m_3\end{array}\right) \tilde a_{l_1m_1} \tilde a_{l_2 m_2} \tilde a_{l_3 m_3}\nonumber
\end{eqnarray}
where 
\begin{eqnarray}
\tilde a_{lm}=\int d\Omega\;Y^{*}_{lm}(\hat{\mathbf n})\,T(\hat{\mathbf n})\,W(\hat{\mathbf n}),
\end{eqnarray}
with $W(\hat{\mathbf n})$ being a foreground mask function.
For the temperature-only case, the cut sky induces mostly a monopole contribution outside the mask, which can be made small by subtracting the monopole term outside the mask \cite{WMAP_fnl,Yadav_fnl_review}. 
To a good approximation, the CMB bispectrum estimated from a cut sky is related to that of a whole sky as follows: \cite{Komatsu_thesis}:
\begin{eqnarray*}
\langle \tilde B^{\mathrm{obs}}_{l_1 l_2 l_3} \rangle\approx f_{\mathrm{sky}}\,\langle B^{\mathrm{obs}}_{l_1 l_2 l_3} \rangle
\end{eqnarray*}
where $f_{\mathrm{sky}}$ is the sky fraction. 

In the weak non-Gaussian limit, the covariance matrix of $\tilde B^{\mathrm{obs}}_{l_1 l_2 l_3}$
is nearly diagonal and its diagonal elements are 
\cite{Komatsu_fnl_review}:
\begin{eqnarray}
\langle (\Delta \tilde B^{\mathrm{obs}}_{l_1 l_2 l_3})^2 \rangle\approx f^{3}_{\mathrm{sky}}(C_{l_1}+N_{l_1})(C_{l_2}+N_{l_2})(C_{l_3}+N_{l_3})\Delta_{l_1 l_2 l_3}\label{B_var}
\end{eqnarray}
where $C_l$ and $N_l$ are the CMB and noise power spectrum, and 
$\Delta_{l_1 l_2 l_3}$ is 1, 2 and 6 for all different three $l_i$, two equal $l_i$ and $l_1=l_2=l_3$ respectively.
In the WMAP noise model, the pixel covariance is given by:
\begin{eqnarray}
\mathbf N_{ij}=\frac{\sigma^2_0}{N_{\mathrm{obs}}(\hat {\mathbf n})} \delta_{ij},\label{N_ij}
\end{eqnarray}
where $\sigma_0$ is the rms noise per observation, and $N_{\mathrm{obs}}(\hat {\mathbf n})$ is the number of observations for the $i$th pixel of the sky direction $\hat {\mathbf n}$. 
Under a white noise assumption, one can estimate noise power spectrum $N_l$ as follows \cite{WMAP5:Cosmology}:
\begin{eqnarray}
N_l=\Omega_{\mathrm{pix}}\int \frac{d\Omega}{4\pi\,f_{\mathrm{sky}}} \frac{\sigma^2_0\,W(\hat {\mathbf n})}{N_{\mathrm{obs}}(\hat {\mathbf n})}, 
\end{eqnarray}
where $\Omega_{\mathrm{pix}}$ is the solid angle per pixel. 
In deriving Eq. \ref{B_var}, we have assumed that the spherical harmonic coefficient of the instrument noise $\eta_{lm}$ has a diagonal covariance (i.e. $\langle \eta_{lm} \eta^*_{l'm'} \rangle = \delta_{ll'} \delta_{mm'} N_l$), which is true only for the case of homogeneous noise and a whole sky coverage.
\begin{figure}[htb!]
\centering\includegraphics[scale=.445]{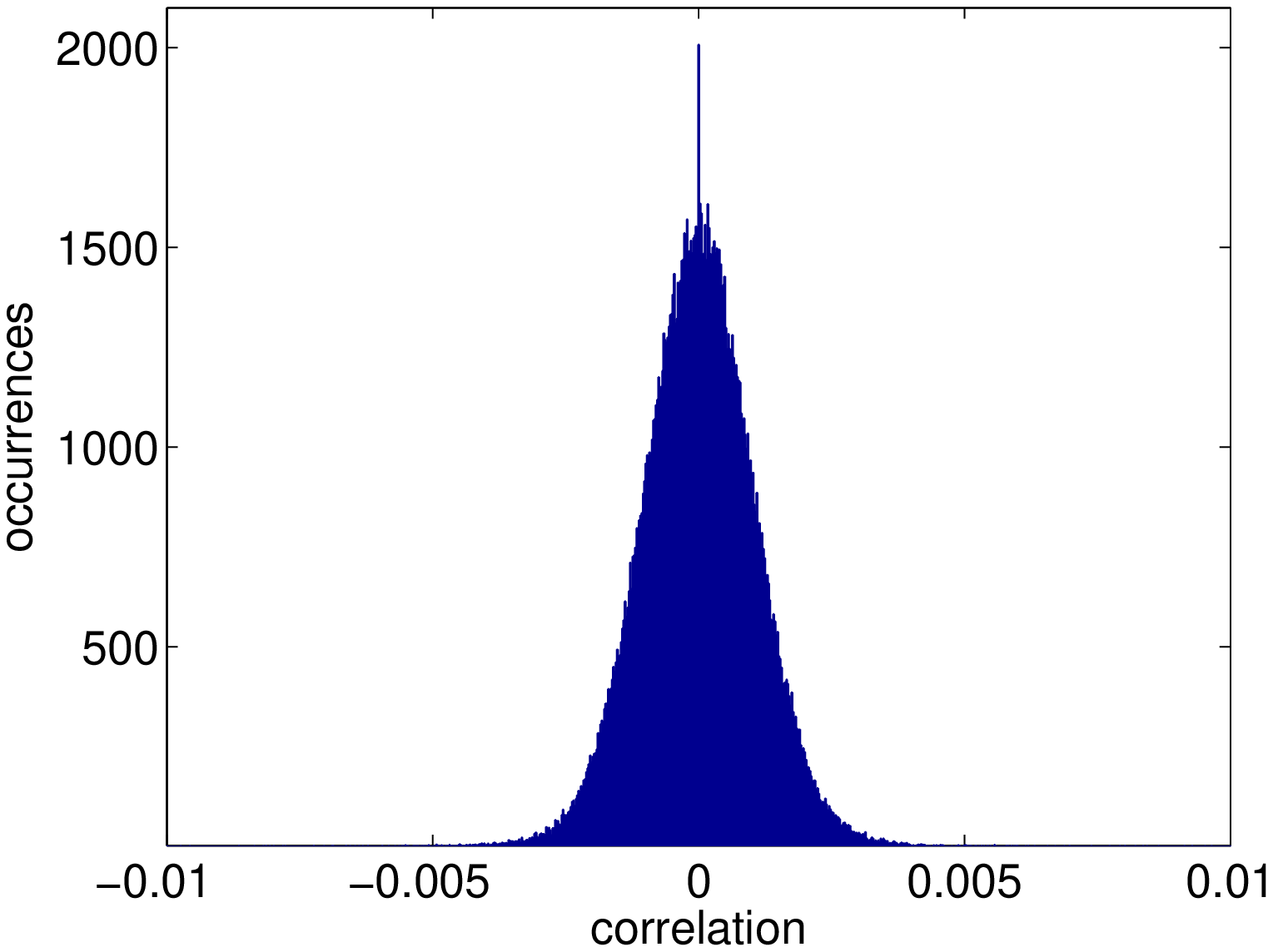}
\centering\includegraphics[scale=.445]{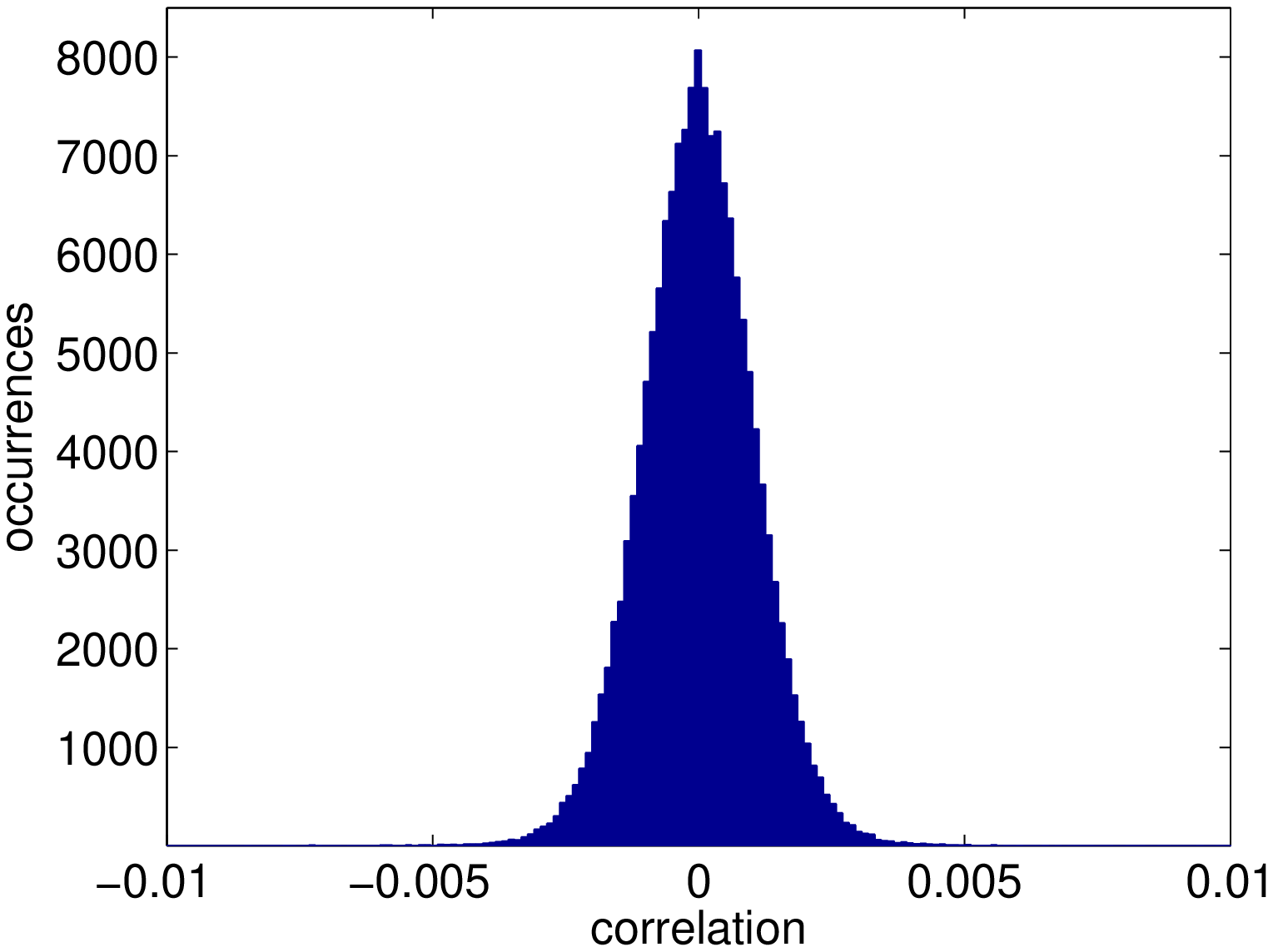}
\caption{Noise correlation $\langle \eta_{lm} \eta^*_{lm'} \rangle/\langle \eta_{lm} \eta^*_{lm} \rangle$ for $l=400$: real part (left) and imaginary part(right), estimated from $5\times10^5$ simulation of inhomogeneous noise on a cut sky.}
\label{noise}
\end{figure}
However, Monte-Carlo simulations show that non-diagonal parts of $\langle \eta_{lm} \eta^*_{l'm'} \rangle$ tend to be less than $1\%$ of the diagonal parts. To be specific, we have simulated $5\times10^5$ noise maps for the V and W band respectively by the WMAP noise model (c.f. Eq. \ref{N_ij}), and applied the foreground mask KQ75 to them.
From the cut-sky simulations, we have estimated the noise covariance matrix $\langle \eta_{lm} \eta^*_{l'm'} \rangle$.
The magnitude of noise correlation turns out to decrease with the relative distance in spherical harmonic numbers (i.e. $|l-l'|$ and $|m-m'|$).
In Fig. \ref{noise}, we show the histogram of noise correlation for the fixed multipole $l=l'=400$ and $m\ne m'$. 
Note that the values of the $y$ axis denote the number of $m$ and $m'$ pairs, whose correlations correspond to the values of the $x$ axis. 
As shown in Fig. \ref{noise}, we find that non-diagonal parts of $\langle \eta_{lm} \eta^*_{l'm'} \rangle$ tend to be less than $1\%$ of the diagonal parts.
Besides, we shall be estimating the bispectrum in the multipole range $l\le 400$, where the Signal-to-Noise Ratio is high (e.g. $17\lesssim C_l/N_l \lesssim 42$ at $l\le 400$) \cite{WMAP7:powerspectra}.
Therefore, we find Eq. \ref{B_var} provides good approximation for the variance.

Point sources, whose fluxes are below a certain detection threshold, are not properly masked out, and therefore require an accurate assessment of their contribution to the bispectrum. 
Assuming a Poisson distribution for the point sources \cite{CMB_src}, their contribution is given by \cite{Komatsu_thesis,WMAP5:Cosmology,Komatsu_fnl_review}:
\begin{eqnarray}
b_{\mathrm{src}}\,I_{l_1 l_2 l_3}, 
\end{eqnarray}
where $b_{\mathrm{src}}$ is an amplitude to be determined, and $I_{l_1 l_2 l_3}$ is given by Eq. \ref{Illl}.

\section{Likelihood function}
\label{ML}
As shown in Eq. \ref{B_obs}, CMB bispectrum observables $B^{\mathrm{obs}}_{l_1 l_2 l_3}$ contain a sum of a large number of terms.
In the weak non-Gaussian limit, the correlation between the terms is insignificant:
\begin{eqnarray}
\frac{\langle \left(\begin{array}{ccc} l_1& l_2 & l_3 \\m_1& m_2& m_3\end{array}\right) \left(\begin{array}{ccc} l_1& l_2 & l_3 \\m'_1& m'_2& m'_3\end{array}\right) a_{l_1m_1} a_{l_2 m_2} a_{l_3 m_3}\, (a_{l_1m'_1} a_{l_2 m'_2} a_{l_3 m'_3})^*\rangle}{
(C_{l_1}+N_{l_1})(C_{l_2}+N_{l_2})(C_{l_3}+N_{l_3})\Delta_{l_1 l_2 l_3}}\approx0,
\end{eqnarray}
where we referred to Eq. \ref{B_var}.
Accordingly, the distribution of $B^{\mathrm{obs}}_{l_1 l_2 l_3}$, which contains a large number of weakly correlated random variables, tends toward a Gaussian function by the central limit theorem. It is worth to note that, unlike CMB power spectrum observables, bispectrum observables $B^{\mathrm{obs}}_{l_1 l_2 l_3}$ contain a sufficiently large number of degree of freedom (e.g. 125) even for the lowest multipole (e.g. $l=2$).
Therefore, we may approximate the likelihood function of the CMB bispectrum as follows: 
\begin{eqnarray}
\mathcal{L}( B_{l_1 l_2 l_3}(\lambda_{\alpha})|B^{\mathrm{obs}}_{l_1 l_2 l_3})=\frac{1}{(2\pi)^{\frac{n}{2}}{\left|\mathbf C\right|}^{\frac{1}{2}}}\exp[-\frac{1}{2}(\mathbf B^{\mathrm{obs}}-\mathbf {B}(\lambda_{\alpha}))^{\dagger}\,{\mathbf C}^{-1}\,(\mathbf B^{\mathrm{obs}}-\mathbf {B}(\lambda_{\alpha}))],\label{likelihood}
\end{eqnarray}
where $\mathbf C$ is a $n\times n$ covariance matrix, $\lambda_\alpha$ are the parameters: $\{f^{\mathrm{local}}_{\mathrm{NL}},f^{\mathrm{equil}}_{\mathrm{NL}},b_{\mathrm{src}}\}$ + cosmological parameters, and $\mathbf B^{\mathrm{obs}}$ and $\mathbf B(\lambda_{\alpha})$ are vectors consisting of the observed and the theoretical bispectrum respectively. 
The theoretical bispectrum $\mathbf B(\lambda_{\alpha})$ is given by:
\begin{eqnarray}
B_{l_1 l_2 l_3}(\lambda_{\alpha})=w_{l_1} w_{l_2} w_{l_3}\,f_{\mathrm{sky}} [f^{\mathrm{local}}_{\mathrm{NL}}\,B^{\mathrm{local}}_{l_1 l_2 l_3}(\lambda_{\alpha})+f^{\mathrm{equil}}_{\mathrm{NL}}\,B^{\mathrm{equil}}_{l_1 l_2 l_3}+b_{\mathrm{src}}\,I_{l_1 l_2 l_3}],\label{B_theory}
\end{eqnarray}
where $w_l$ is the effective window function, which is the product of the beam transfer function and the pixel window function.

\begin{figure}[htb!]
\centering\includegraphics[scale=.44]{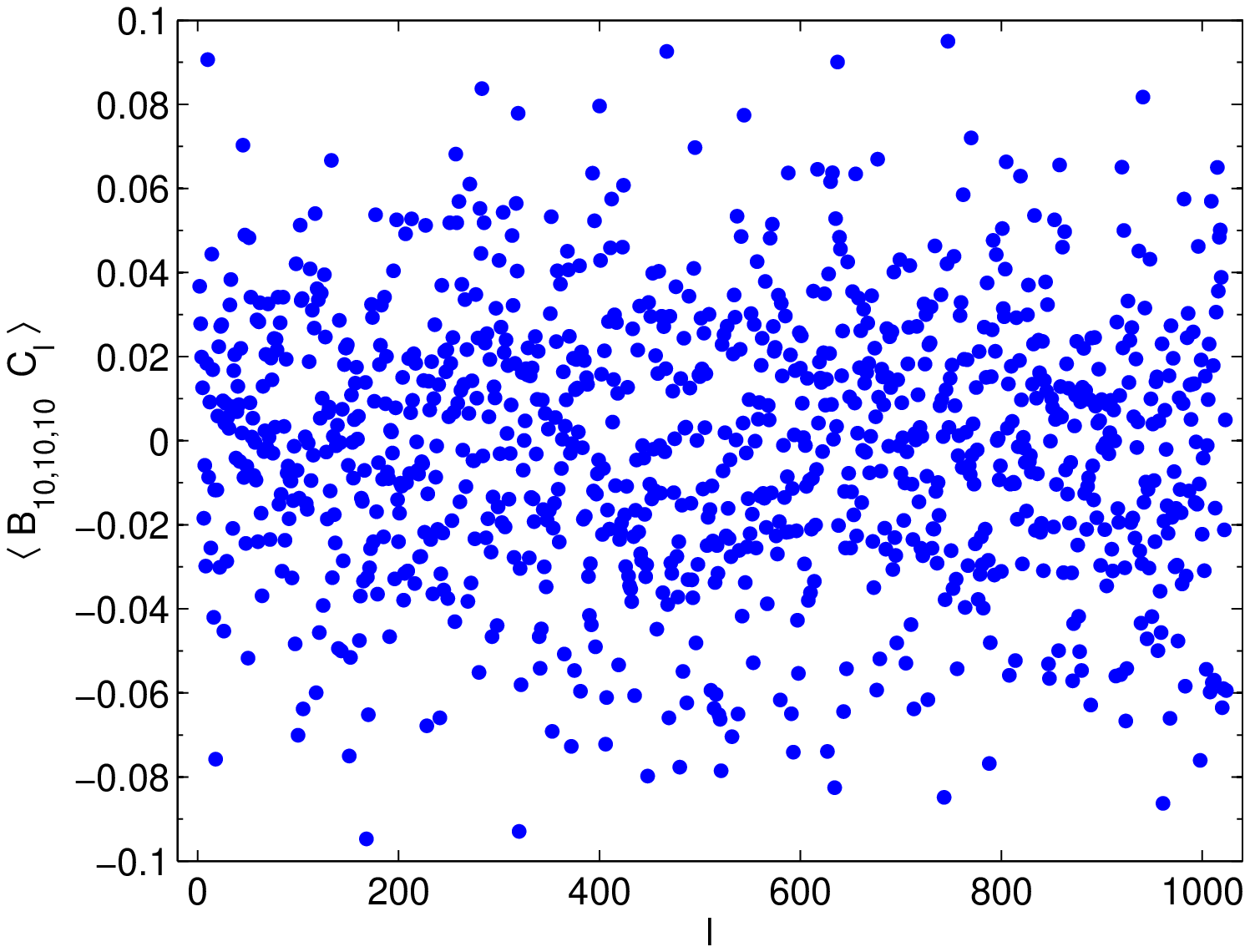}
\centering\includegraphics[scale=.44]{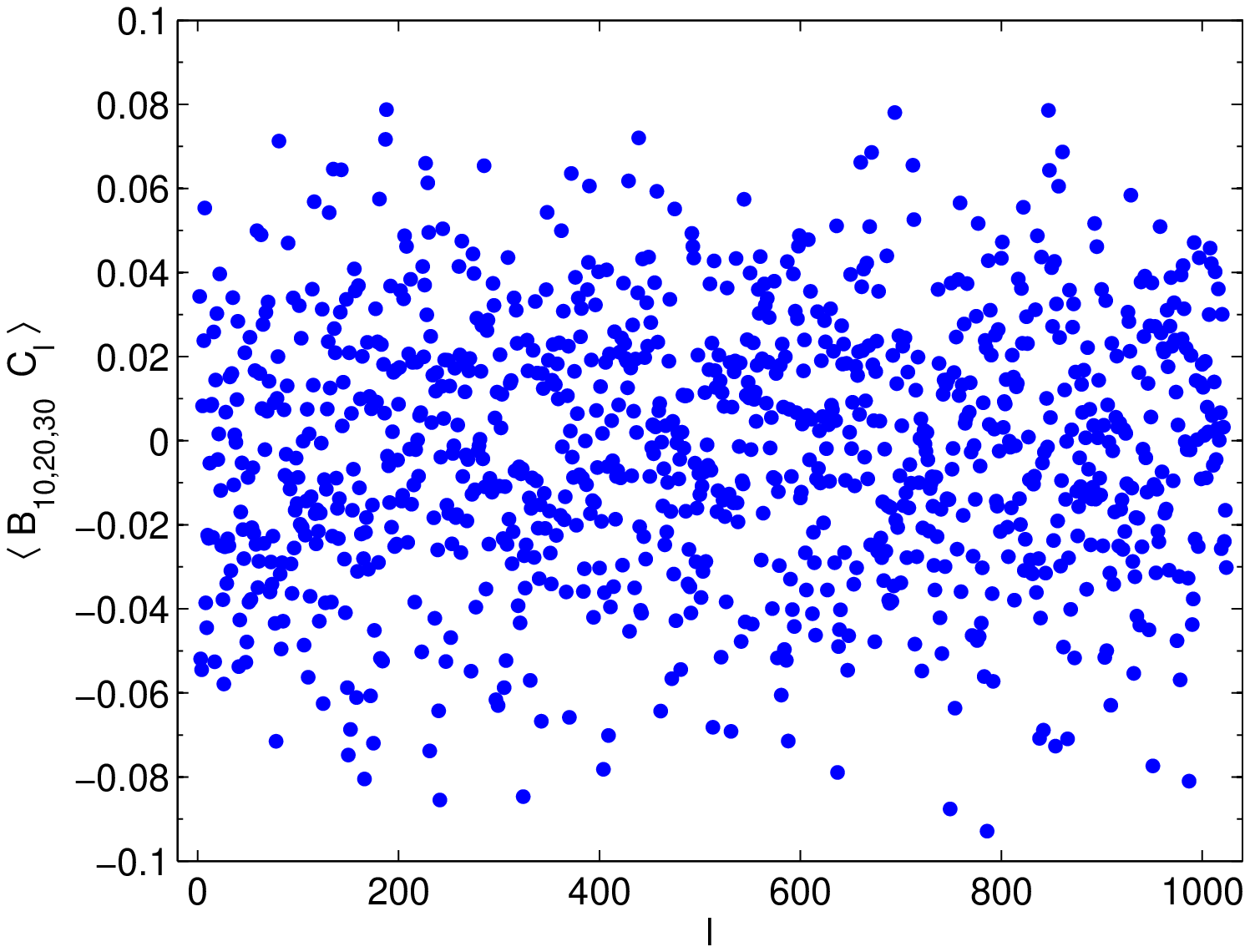}
\centering\includegraphics[scale=.44]{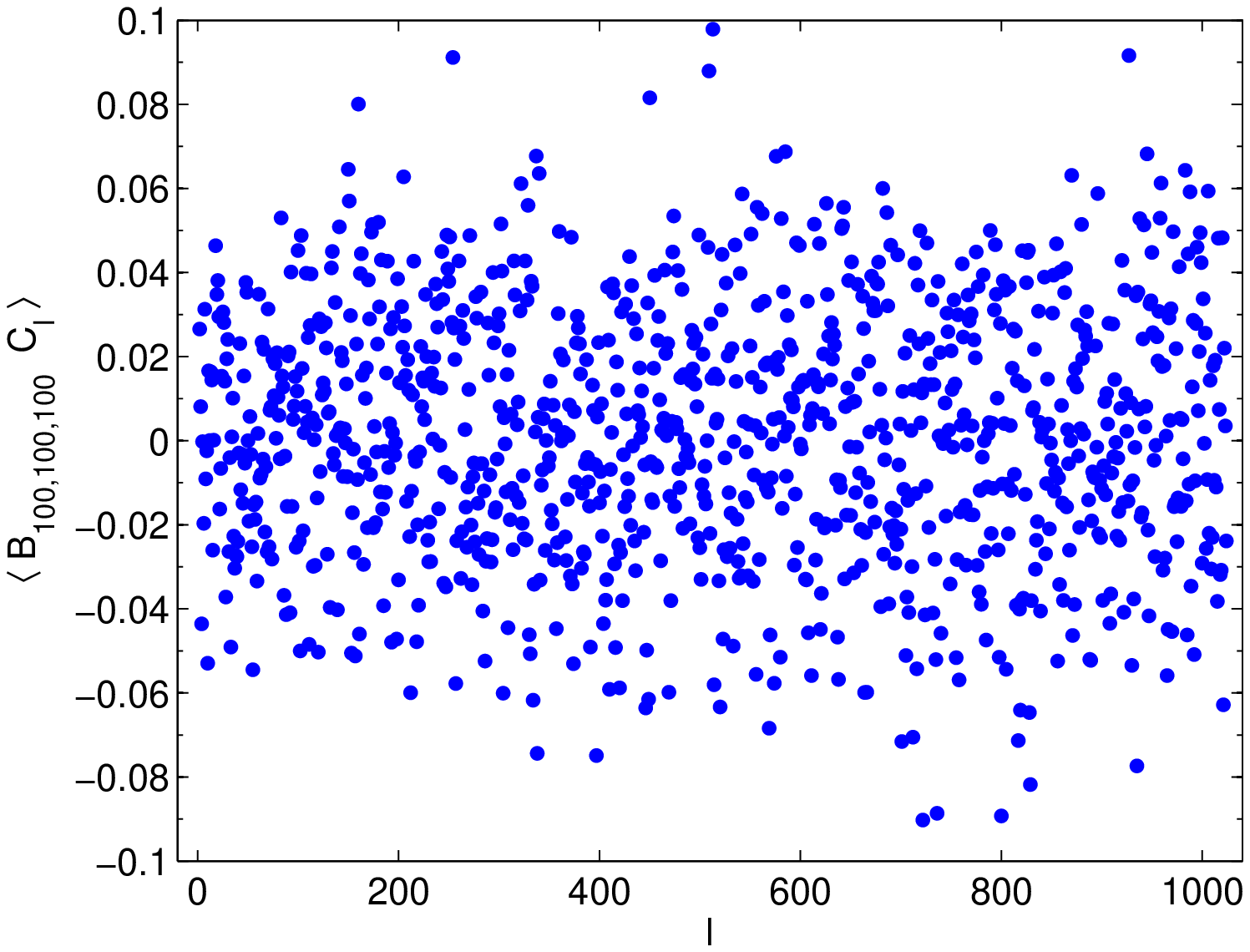}
\centering\includegraphics[scale=.44]{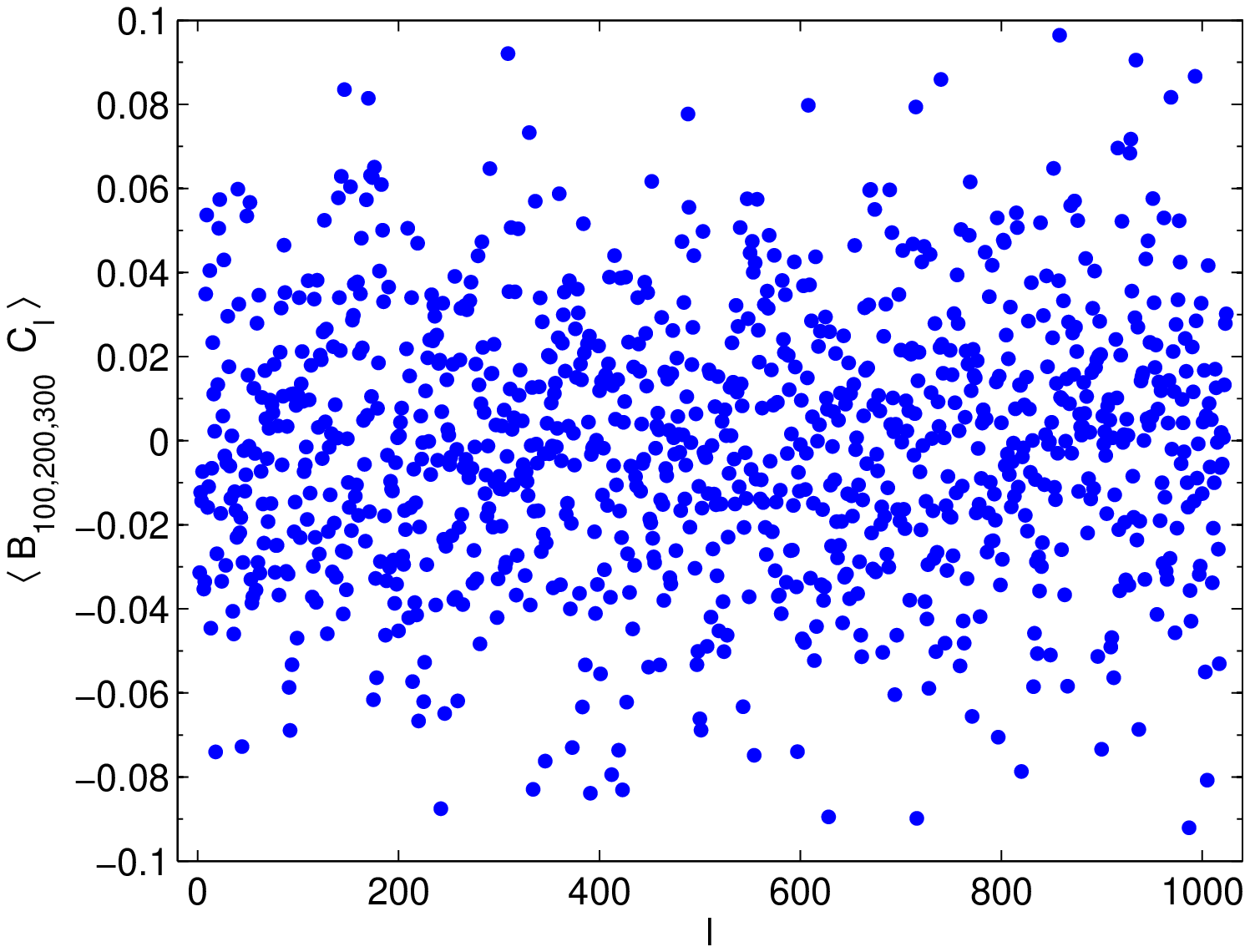}
\centering\includegraphics[scale=.44]{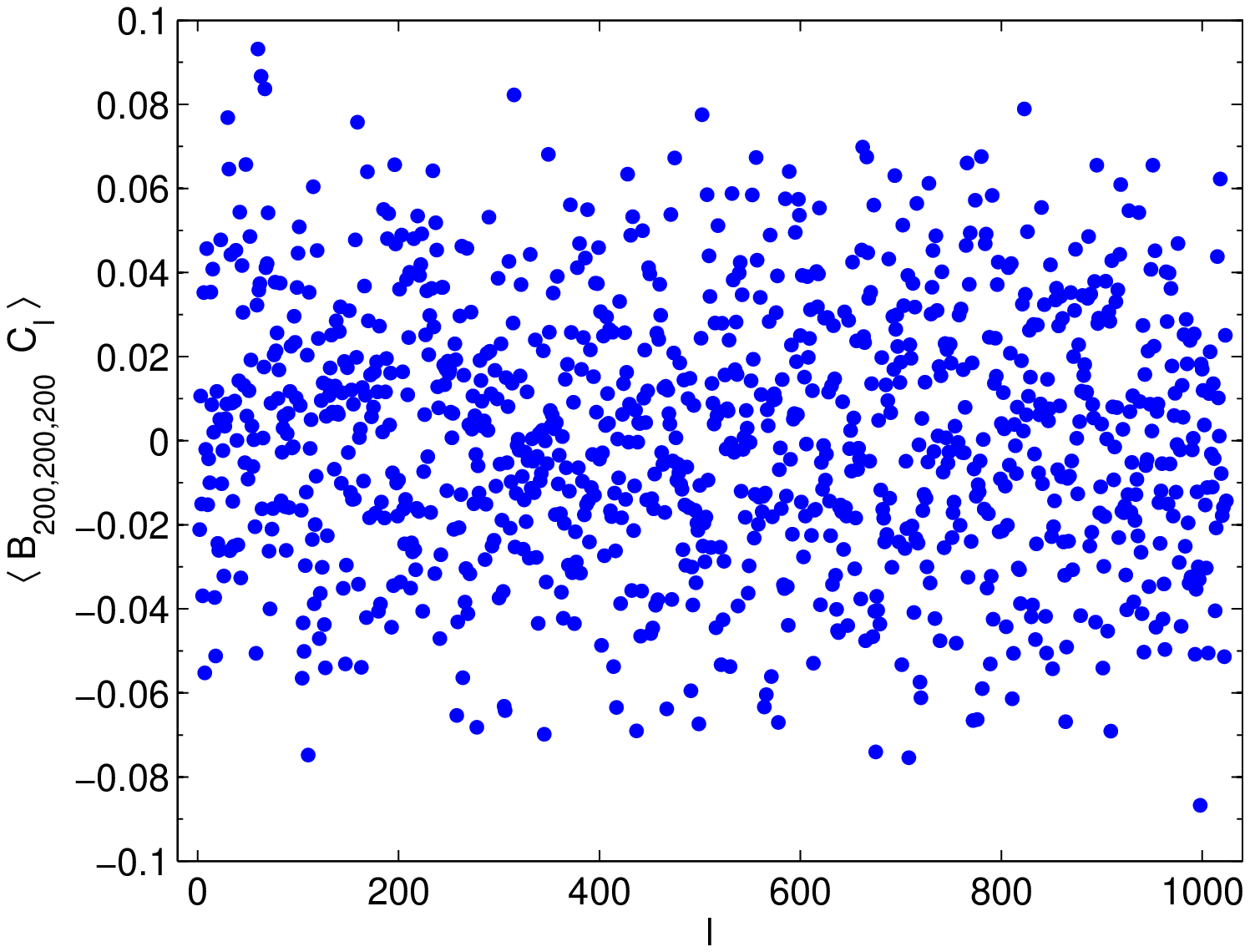}
\centering\includegraphics[scale=.44]{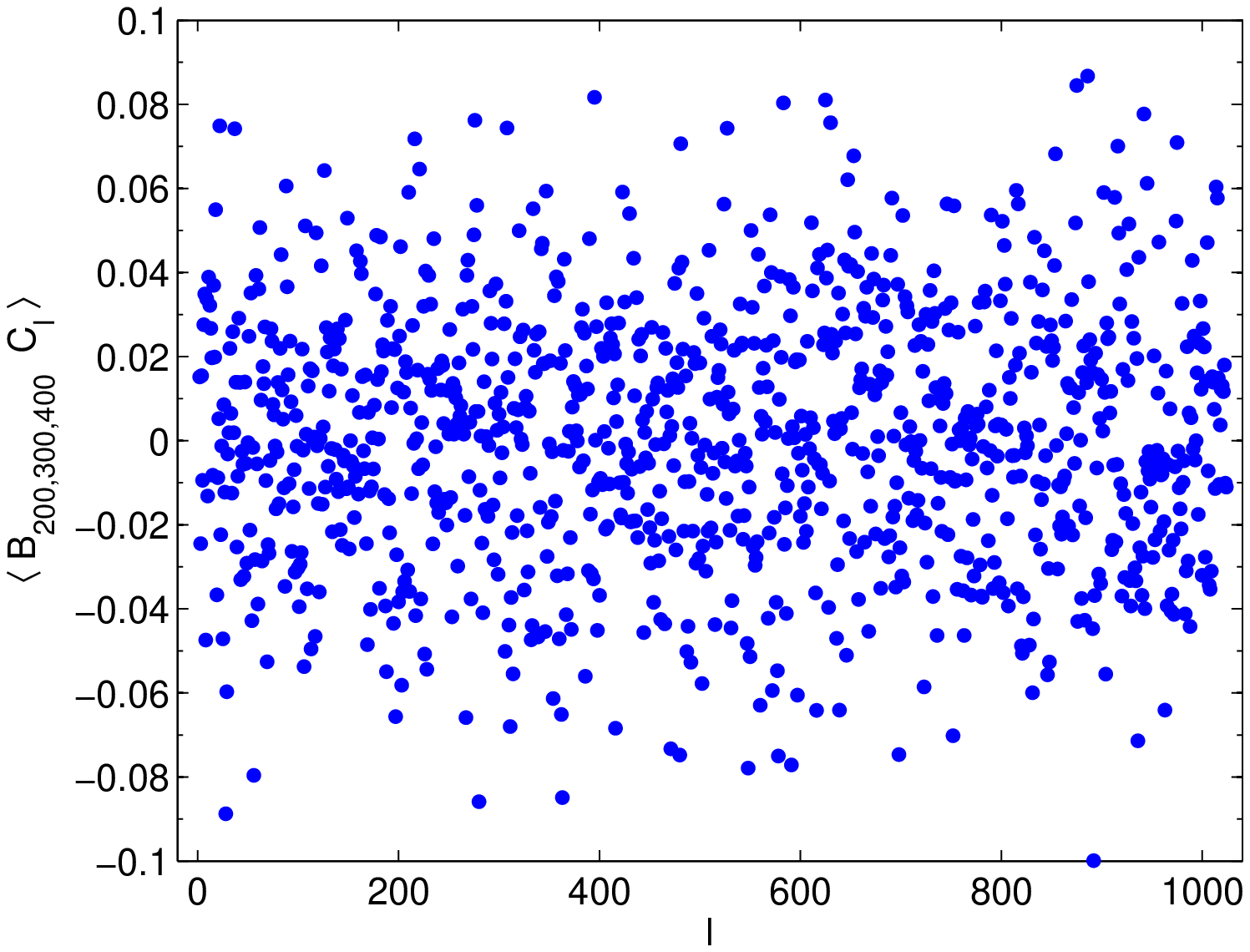}
\caption{Correlation between CMB power spectrum and bispectrum: $\langle B_{l_1,l_2,l_3}\:C_{l}\rangle$}
\label{correlation}
\end{figure}
In order to derive a total likelihood, we have investigated the correlation between CMB power spectrum and bispectrum.
Due to the complexity of analytic derivation, we have made numerical investigation, using simulations produced by \cite{fnl_sim}. 
The simulations are provided in such a way that we may set the value of $f^{\mathrm{local}}_{\mathrm{NL}}$.
In Fig. \ref{correlation}, we show the correlation estimated from 1000 simulations, where $f^{\mathrm{local}}_{\mathrm{NL}}$ is set to 40. 
As shown in Fig. \ref{correlation}, we find the correlation between the CMB power spectrum and bispectrum is insignificant.
Therefore, to a good approximation, we may construct a full likelihood function as follows:
\begin{eqnarray}
\mathcal{L}(C_l|C^{\mathrm{obs}}_l)\times \mathcal{L}( B_{l_1 l_2 l_3}(\lambda_{\alpha})|B^{\mathrm{obs}}_{l_1 l_2 l_3}), 
\end{eqnarray}
where $\mathcal{L}(\bar C_l|C_l)$ is the likelihood function associated with CMB power spectrum, whose approximate and optimal expression is found in \cite{WMAP1:parameter_method}. 
The corresponding log-likelihood is given by:
\begin{eqnarray}
\log\mathcal{L}(C_l|C^{\mathrm{obs}}_l) + \log \mathcal{L}( B_{l_1 l_2 l_3}(\lambda_{\alpha})|B^{\mathrm{obs}}_{l_1 l_2 l_3}) \label{full_likelihood}.
\end{eqnarray}
The log-likelihood of the WMAP CMB power spectrum, which corresponds to the first term in Eq. \ref{full_likelihood}, is provided by the WMAP team's code and well-integrated into the widely used Monte-Carlo sampling code (i.e. \texttt{CosmoMC}). Therefore, we are going to be mainly concerned with adding the second term of Eq. \ref{full_likelihood} to the existing code. 

As shown in Eq. \ref{full_likelihood}, the likelihood function has dependency on several cosmological parameters. 
Therefore, we are going to explore multi-dimensional parameter space via MCMC sampling.
In this way, we are going to simultaneously fit all cosmological parameters, including $f_{\mathrm{NL}}$.

\section{Analysis of the WMAP data}
\label{analysis}
We have co-added WMAP 7 year foreground-reduced maps of the V band and the W band, and applied the KQ75 mask to the co-added map \cite{WMAP5:fg,WMAP7:fg,WMAP7:basic_result}.
From the masked V+W map, we have estimated $B^{\mathrm{obs}}_{l_1 l_2 l_3}$ up to $l\le 400$ via Eq. \ref{B_obs}, where the Wigner 3j symbols have been computed efficiently using the recurrence relation \cite{Wigner3j}. It took several hours to compute $B^{\mathrm{obs}}_{l_1 l_2 l_3}$ ($l\le 400$), which is, however, a manageable one-time cost. 
\begin{figure}[htb!]
\centering\includegraphics[scale=.73]{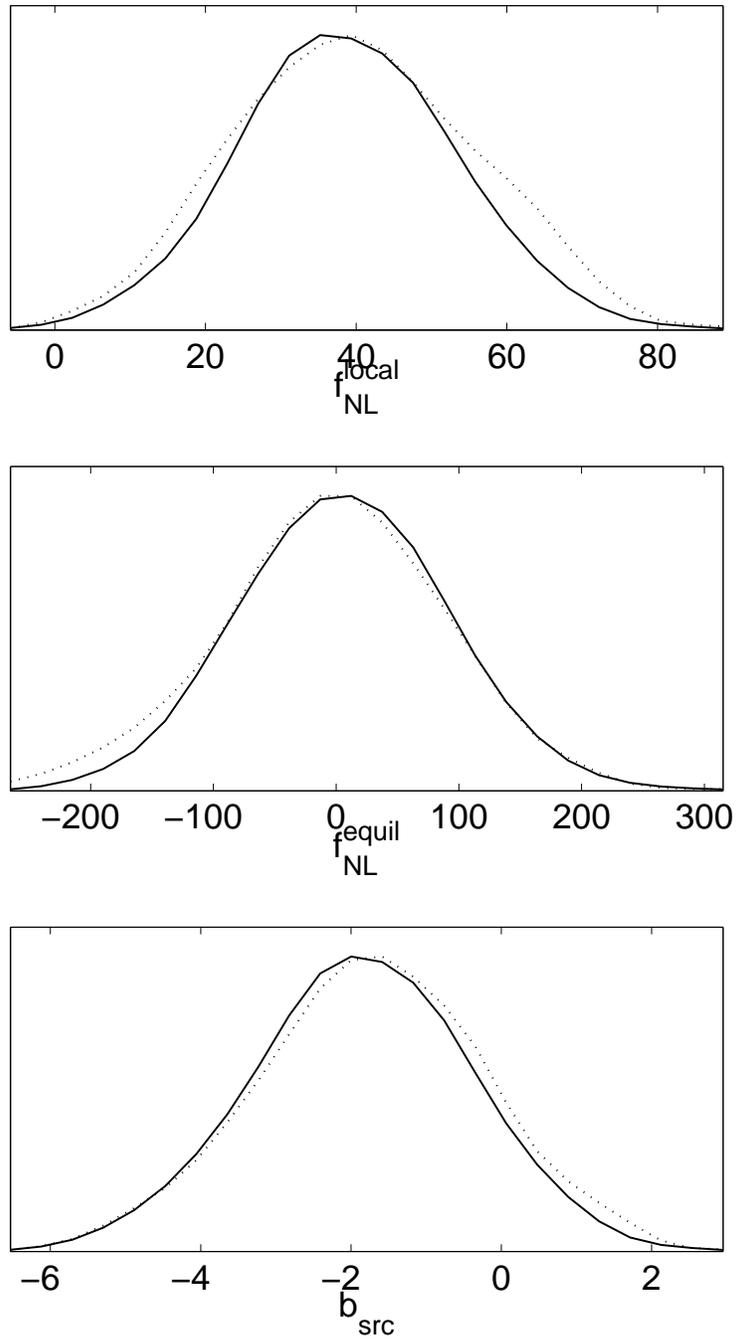}
\caption{The likelihoods of $f_{\mathrm{NL}}$ and $b_{\mathrm{src}}$: solid (dotted) lines denote the marginalized (mean) likelihood.}
\label{like1}
\end{figure}
\begin{figure}[htb!]
\centering\includegraphics[scale=0.6]{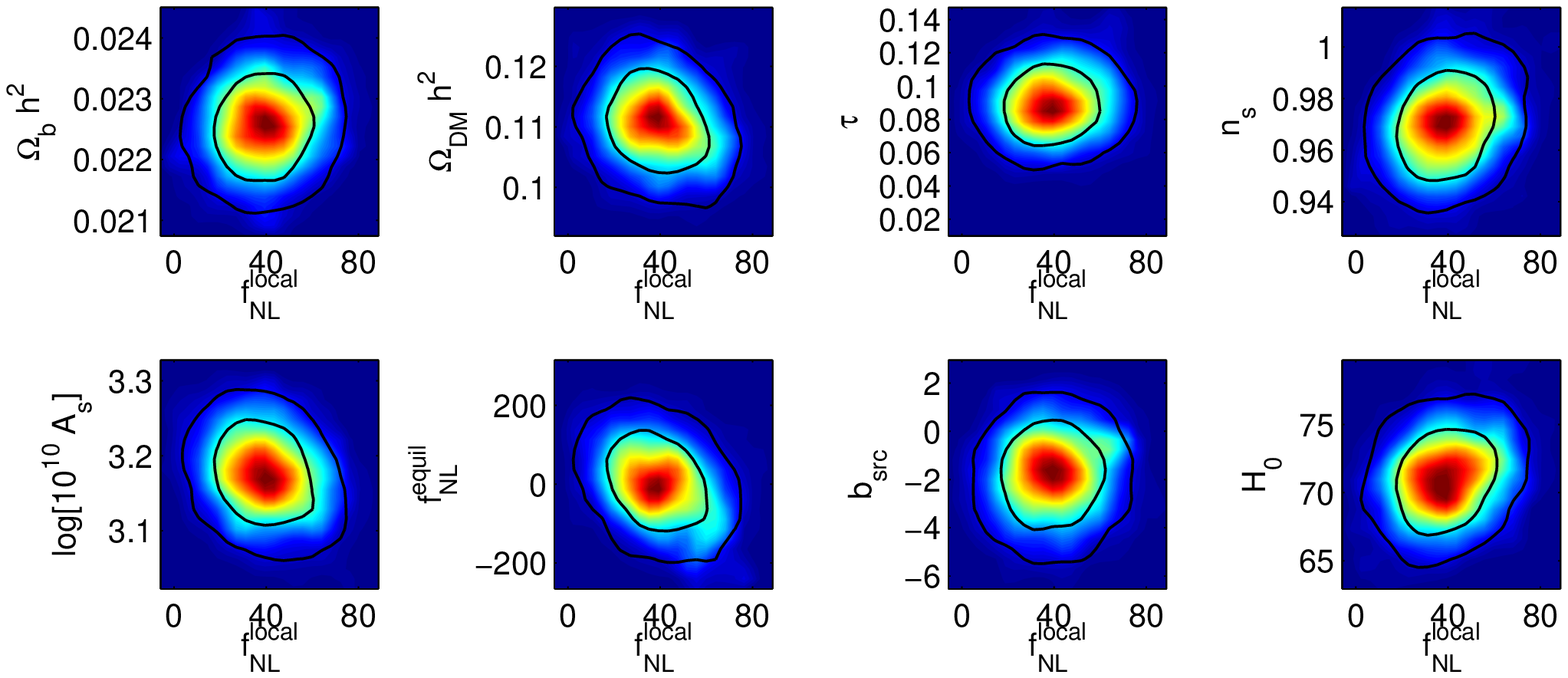}
\centering\includegraphics[scale=0.6]{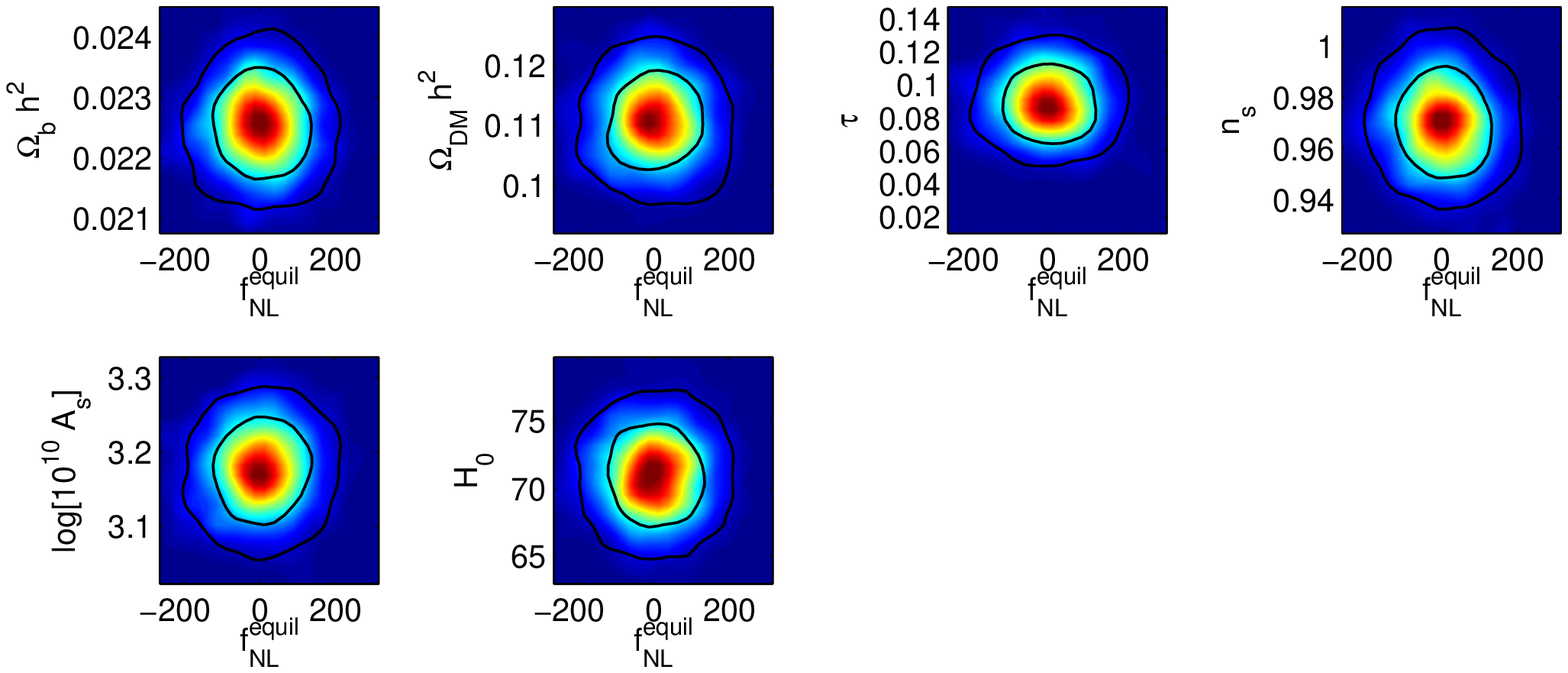}
\centering\includegraphics[scale=0.6]{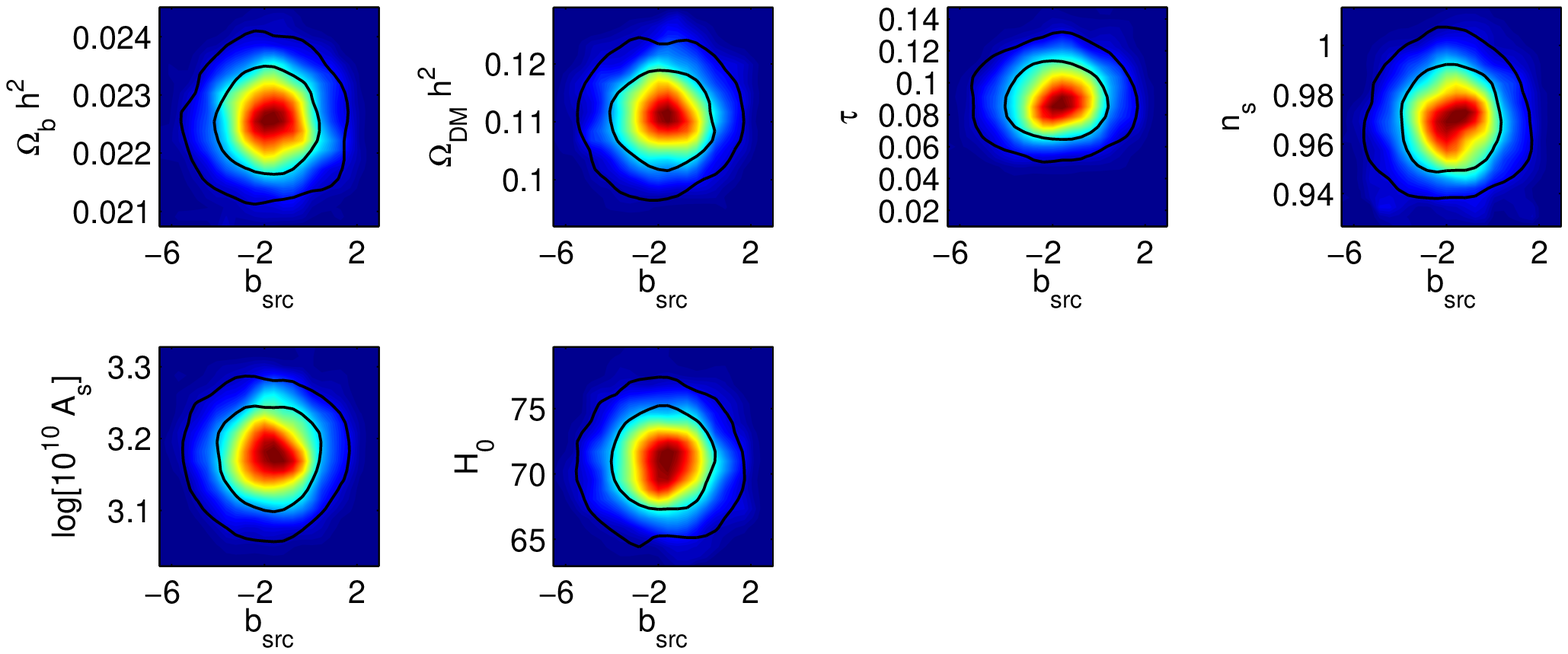}
\caption{Marginalized likelihoods in the plane of $f_{\mathrm{NL}}$ versus cosmological parameters: the contour lines denote 1 and 2 $\sigma$ intervals.}
\label{like2}
\end{figure}
We modified the \texttt{CosmoMC} package so that we can include the bispectrum likelihood (i.e. Eq. \ref{likelihood}) and parameters ($\{f^{\mathrm{local}}_{\mathrm{NL}},f^{\mathrm{equil}}_{\mathrm{NL}},b_{\mathrm{src}}\}$) in the Markov Chain Monte Carlo (MCMC) analysis \cite{CAMB,CosmoMC}. Note that the power spectrum likelihood is provided by the WMAP team and is already integrated into the \texttt{CosmoMC} package.
In order to avoid unnecessary computations, we have added $f_{\mathrm{NL}}$ and $b_{\mathrm{src}}$ as amplitude parameters so that
a radiation transfer function (i.e. $g_l(k)$ in Eq. \ref{alm}) or the normalized bispectrum (i.e. Eq. \ref{B_local} and \ref{B_equil})
is not re-computed for a change of the parameters. For a cosmological model, we have considered $\Lambda$CDM + SZ effect + weak-lensing, where the cosmological parameters are
$\lambda \in \{\Omega_b,\Omega_c,\tau,n_s, A_s, A_{sz}, H_0, f^{\mathrm{local}}_{\mathrm{NL}},f^{\mathrm{equil}}_{\mathrm{NL}},b_{\mathrm{src}}\}$. 
For data constraints, we have used the WMAP 7 year power spectrum and the bispectrum data described above. By running the modified \texttt{CosmoMC} package \cite{CAMB,CosmoMC} on an MPI cluster with 6 chains, we have explored the likelihood in the multi-dimensional parameter space. For a convergence criterion, we have adopted Gelman and Rubin's ``variance of chain means'' and set the R-1 statistic to $0.03$ for a stopping criterion \cite{Gelman:inference,Gelman:R1}.  
Running a single CPU for each MPI process (total 6 CPUs), we reached the target convergence in less than a day.
Once we reached the target convergence, we analyzed the MCMC samples with the \texttt{CosmoMC} package and obtained the parameter likelihood of $f_{\mathrm{NL}}$. 
In Fig.~\ref{like1}, we show the marginalized and mean likelihoods of $f_{\mathrm{NL}}$ and $b_{\mathrm{src}}$. The marginalized distribution shows the probability in the reduced dimension of parameter space and the mean likelihood is associated with the best-fit value \cite{CosmoMC}. For a Gaussian distribution, marginalized and mean likelihoods are identical. A shown in Fig.~\ref{like1}, there exist slight discrepancies between the marginalized and mean likelihoods, which indicate the deviation of the parameter likelihood from a Gaussian shape.
\begin{figure}[htb!]
\centering\includegraphics[scale=.65]{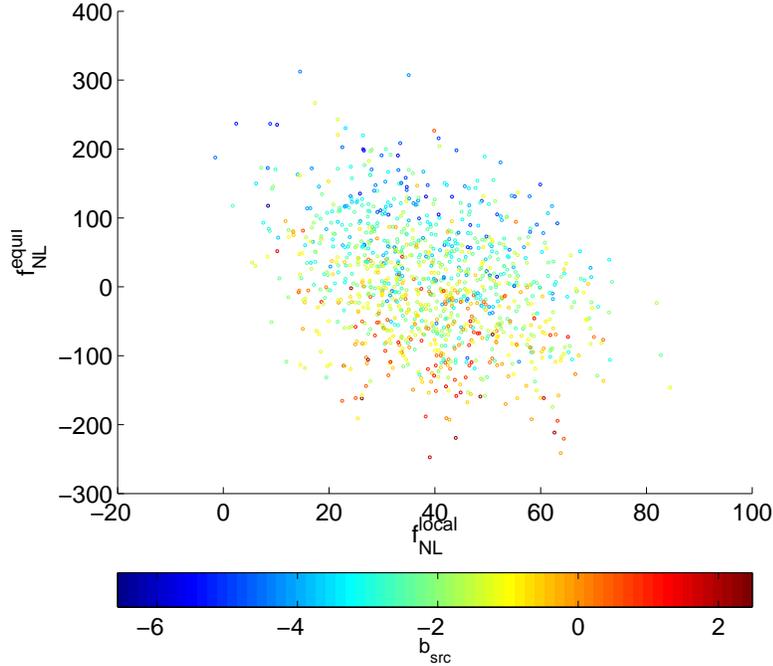}
\caption{MCMC samples: the $x$ and $y$ coordinates correspond to the values of $f^{\mathrm{local}}_{\mathrm{NL}}$ and $f^{\mathrm{equil}}_{\mathrm{NL}}$, color-coded according to the value of $b_{\mathrm{src}}$, marginalized over cosmological parameters.}
\label{like3}
\end{figure}
\begin{table}[htb!]
\centering
\caption{The best-fit values and 68\% confindence intervals for $f_{\mathrm{NL}}$ and the cosmological parameters: ($\Lambda$CDM + sz + lens)}
\begin{tabular}{cc}
\hline\hline 
parameter &  estimate\\
\hline
$f^{\mathrm{local}}_{\mathrm{NL}}$ & $39^{+13}_{-14}$\\
$f^{\mathrm{equil}}_{\mathrm{NL}}$ & $-76<f^{\mathrm{equil}}_{\mathrm{NL}}<88$\\
$b_{\mathrm{src}}$ [$10^{-5} \mu\mathrm{K}^3\,\mathrm{sr}^2]$ & $-2.4^{+2}_{-0.9}$ \\
$\Omega_{b}\,h^2$  &  $0.0225^{+0.0007}_{-0.0005}$\\ 
$\Omega_{c}\,h^2$  &  $0.113^{+0.003}_{-0.008}$ \\ 
$\tau$  & $0.083^{+0.013}_{-0.002}$\\ 
$n_s$ & $0.967^{+0.017}_{-0.011}$ \\ 
$\log[10^{10}A_s]$  &  $3.183^{+0.037}_{-0.06}$ \\ 
$A_{\mathrm{sz}}$  &  $1.42^{+0.58}_{-1.42}$ \\ 
$H_0$  [km/s/Mpc]&  $70.35^{+3.19}_{-1.8}$ \\ 
\hline
\end{tabular}
\label{parameters1}
\end{table}
In Fig.~\ref{like2}, we have plotted the marginalized likelihoods in the plane of $f_{\mathrm{NL}}$ versus other parameters.
As shown in Fig.~\ref{like2}, there exist non-negligible deviations of the parameter likelihoods from a Gaussian shape.
To be specific, the shape of parameter likelihood (e.g. contours of the same color) are poorly fitted by a Gaussian function.
In Fig.~\ref{like3}, we show the MCMC samples, where the $x$ and $y$ coordinates correspond to the values of $f^{\mathrm{local}}_{\mathrm{NL}}$ and $f^{\mathrm{equil}}_{\mathrm{NL}}$, and the value of $b_{\mathrm{src}}$ is color-coded. Note that MCMC samples shown in Fig.~\ref{like3} are obtained after marginalized over the cosmological parameters.
In Table \ref{parameters1}, we show the best-fit values and 1$\sigma$ intervals for the cosmological parameters as well as for $f_{\mathrm{NL}}$. 
As shown in Fig.~\ref{like1} and Table \ref{parameters1}, our result is in good agreement with the current results on $f_{\mathrm{NL}}$ and the WMAP concordance model \cite{WMAP5:Cosmology,WMAP7:Cosmology}.
However, the confidence interval on $f^{\mathrm{local}}_{\mathrm{NL}}$ is different, which we attribute to a difference in the confidence interval estimation.
While we have estimated the confidence interval by making a full exploration of the parameter likelihood, most of the existing works have relied on a Fisher matrix, which yields accurate results only for a parameter likelihood of a Gaussian shape.

\section{Discussion}
\label{Discussion}
We have made an integrated MCMC analysis of the primordial non-Gaussianity ($f_{\mathrm{NL}}$), using the WMAP bispectrum and power spectrum.
In our analysis, we have simultaneously constrained $f_{\mathrm{NL}}$ and the cosmological parameters so that the uncertainties of the cosmological parameters can properly propagate into the $f_{\mathrm{NL}}$ estimation. 
From the parameter likelihoods deduced from the MCMC samples, we find that the parameter likelihood slightly deviates from a Gaussian shape, which makes the Fisher matrix estimation less accurate. Therefore, we have estimated the confidence intervals by exploring the parameter likelihood without using a Fisher matrix.
We find that our best-fit values agree well with the existing results. However, we find that the confidence interval is slight different, which we attribute to the difference in the confidence estimation. 
Mostly due to the increase in the number of parameters, our approach requires a much higher computational load than the estimator of fixed cosmological parameters.
However, with further optimization, our approach will be also feasible for the high angular resolution Planck data.

\section{Acknowledgments}
We are grateful to the anonymous referee for thorough reading and comments, which greatly helped us to improve the clarity of this paper.
We thank Pavel Naselsky, Hael Collins, Eiichiro Komatsu and Anthony Lewis for helpful discussions. 
We acknowledge the use of the Legacy Archive for Microwave Background Data Analysis (LAMBDA), and the non-Gaussianity simulations produced by \cite{fnl_sim}.
This work made use of HEALPix \cite{HEALPix:Primer,HEALPix:framework} and the \texttt{CosmoMC} package. 
This work is supported in part by Danmarks Grundforskningsfond, which allowed the establishment of the Danish Discovery Center.
This work was supported by FNU grant 272-06-0417, 272-07-0528 and 21-04-0355.

\bibliographystyle{unsrt}
\bibliography{/home/tac/jkim/Documents/bibliography}
\end{document}